\documentclass[twocolumn,showpacs,preprintnumbers,amsmath,amssymb,pre]{revtex4-1}

\usepackage{graphicx}
\usepackage{dcolumn}
\usepackage{overpic}
\usepackage{float} 
\usepackage{bbm}

\begin{document}

\title{Topological  instabilities in ac-driven bosonic systems} 

\author{ G. Engelhardt$^1$    }
\email{georg@itp.tu-berlin.de}

\author{ M. Benito$^2$}

\author{ G. Platero$^2$}

\author{ T. Brandes$^1$}

\affiliation{%
$^1$Institut f\"ur Theoretische Physik, Technische Universit\"at Berlin, Hardenbergstr. 36, 10623 Berlin, Germany \\
$^2$Instituto de Ciencia de Materials de Madrid, CSIC, 28049 Madrid, Spain}%

\begin{abstract}
Under non-equilibrium conditions,  bosonic modes can become dynamically unstable with an exponentially growing  occupation. On the other hand, topological band structures give rise to symmetry protected midgap states. In this letter, we investigate the interplay of instability and topology. Thereby,
 we establish a general relation between topology and instability under ac-driving.
 We apply our findings to create dynamical instabilities which are strongly localized at the boundaries of a finite-size system. As these localized instabilities are protected by symmetry, they can be considered as topological instabilities.
\end{abstract}


\maketitle

\newcommand{\up}{\uparrow}
\newcommand{\down}{\downarrow}
\newcommand{\rAr}{\rightarrow}

\textit{Introduction.}
\label{sec:Introduction}
%
 Due to their underlying symplectic structure, bosonic systems can exhibit so-called dynamical instabilities~\cite{Kawaguchi2012,Goren2007,Arnold1989}. This effect can occur in the presence of non-particle conserving terms which appear, e.g., in the Bogoliubov excitations of Bose-Einstein condensates. Thereby, the bosonic occupation of a mode grows exponentially in time due to a non-equilibrium state of the system. On the other hand, the theory of topological band structures predicts symmetry protected midgap states~\cite{Bernevig2013,Fu2006,Hasan2010,Jotzu2014,Shindou2013,Shindou2013a,Engelhardt2015,Furukawa2015,Bardyn2015,Peano2015}.
 A priory, dynamical instability and topology are independent phenomena.

In this letter, we  formalize a relation between instabilities and topology under ac driving.
More precisely, we demonstrate that different topological phases are always separated by regions of instability.
Using this relation,  we demonstrate how to employ topology to systematically  engineer topologically protected dynamical instabilities. 
 Thereby,  spatially localized midgap modes are rendered dynamically  unstable with exponentially growing bosonic occupation  as has been exemplary proposed for Bose-Einstein condensates in Refs.~\cite{Furukawa2015,Barnett2013,Galilo2015}, and for  photonic systems~\cite{Peano2016}. 
 Here, we suggest a very flexible tool in the form of 
ac-fields in order to engineer topological instabilities governed by 
corresponding artificial, effective Hamiltonians~\cite{Bastidas2012}.
 This simultaneously provides the possibility to detect the midgap  states as their occupation increases exponentially in time.
Topological instabilities are an effect with no direct analogue in fermionic topological insulators.
This stresses the need for a more intensive investigation of topological effects in bosonic systems.

 In fermionic systems, ac-driving has been
applied to control  topological phases~\cite{Rechtsman2013,Wang2013,Oka2009,Kundu2013,Wang2014,Benito2014,Mikami2015}.
 In particular, the topology
of a band can  change if there is a degeneracy of the form $\epsilon_{i'} = \epsilon_i + \Omega$, where $\Omega$
denotes the driving frequency. As this is a single-particle effect, it can also appear  in bosonic systems~\cite{Salerno2016,Fleury}. 
The main challenge, however, is that  dynamical instabilities appearing in bosonic systems generated by an ac-driving constitute an obstruction in the search for a stable system with non-trivial topology~\cite{Salerno2016}:
by slightly changing  parameters, the system  might get unintentionally unstable within the bulk, which obscures the existence of midgap states. For the one-dimensional Hamiltonian under consideration, we show how to employ stability diagrams  to facilitate the search of adequate system parameters. 
The latter can be regarded as a higher-dimensional version of the famous Arnold tongues in parametrically driven oscillators~\cite{Arnold1989}.

Selective enhancement of   edge states and related effects can be achieved using different approaches, e.g.,
non-Hermitian Hamiltonians~\cite{Malzard2015,Schomerus2013,Poli2015,Zeuner2015,Malzard2015}.
 Furthermore, in  a driven spin chain the crossing of a topological phase transitions is accompanied by a Kibble-Zurek scaling phenomenon~\cite{Russomanno2015}. These effects raise the question, to what extend the instability-topology relation established here for bosonic ac-driven systems can be generalized to other fields of physics.

\textit{The system.}
\label{sec:TheSystem}
Bogoliubov Hamiltonians are important in many areas of physics. For instance, they appear by an expansion of Hamiltonians
describing interacting  bosonic particle or polariton condensates in orders of the fluctuations~\cite{Kawaguchi2012,Goren2007,Buecker2011,Barnett2013}. They also describe
excitations in magnonic crystals~\cite{Shindou2013,Shindou2013a} or in quantum-optical systems \cite{Emary2003}. More generally,
they appear in the linear stability analysis of nonlinear bosonic systems. 

\begin{figure}[t]
  \centering
  \includegraphics[width=1\linewidth]{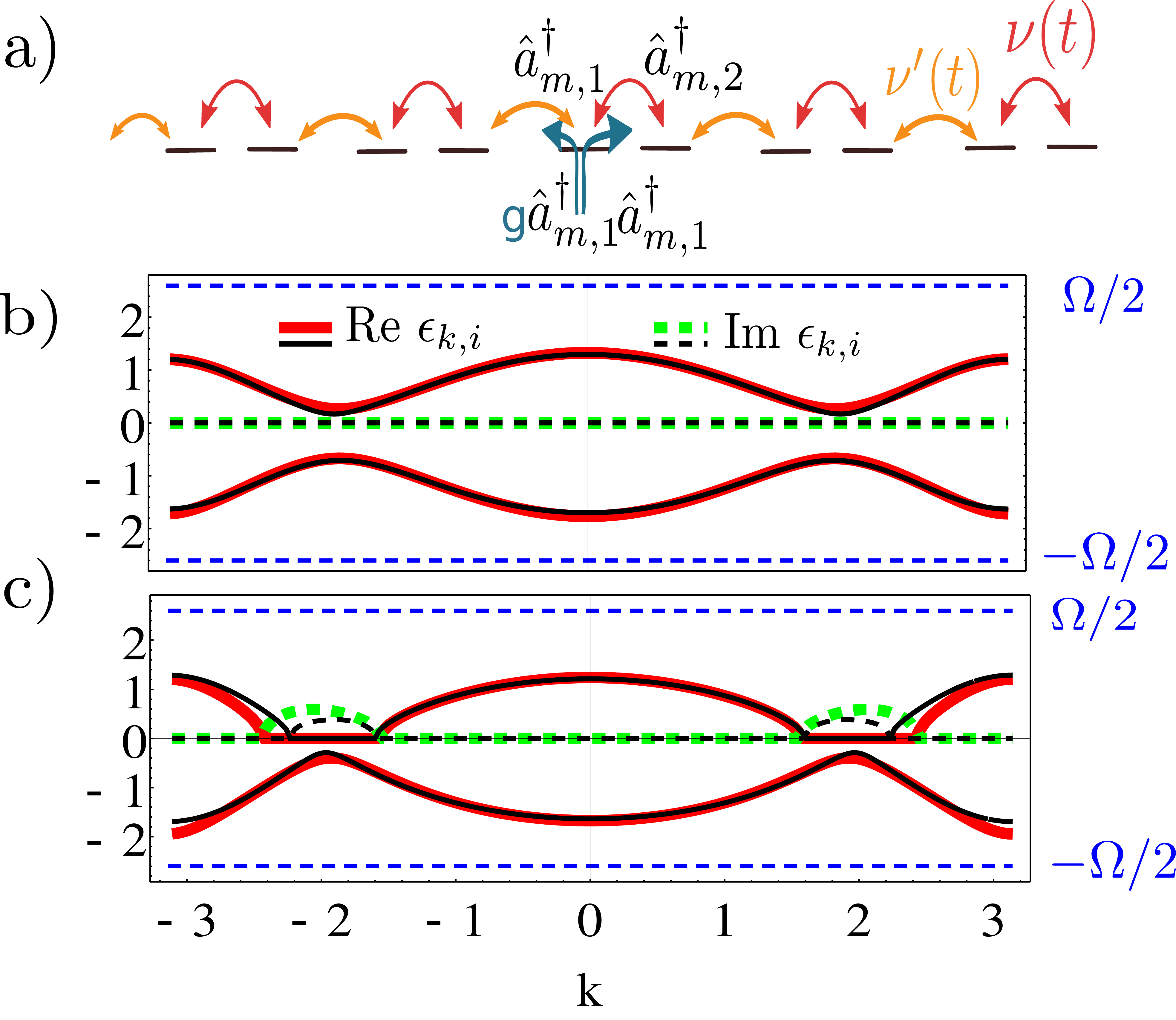}
  \caption{(Color online) \textbf{(a)} Sketch of the system. 
  \textbf{(b)} and \textbf{(c)} depict  quasienergy spectra     with colored lines for  $\nu_0=1.5$, $\nu'_0=0$, $\nu_1=3$, $\mu=-5$ and $\Omega=5.2$. 
  We choose  $\nu'_1=11$ and $\nu'_1=6$ in (b) and (c), respectively.
   Black lines depict the spectrum of an  effective Hamiltonian~\cite{Supplementals}. All quantities are expressed in units of $g$.
  }
  \label{fig:system}
  \end{figure}

We analyze a one-dimensional system of coupled bosonic modes which is subjected to periodic driving.
However, we emphasize that the relation between instability and topology established here apply also for higher-dimensional systems.
A Hamiltonian allowing for a systematic investigation of the topological instabilities which we are interested in reads
 \begin{align}
   H &= \sum_m  - \left(\nu (t)  \hat a_{m,1}^\dagger\hat a_{m,2} + \nu' (t)  \hat a_{m,2}^\dagger\hat a_{m+1,1}    + \text{h.c.}\right)
           \nonumber \\
        &+g \sum_{m,s=\{1,2\}} \left(   \hat a_{m,s}^\dagger\hat a_{m,s}^\dagger   + \text{h.c.} \right) -  \sum_{m,s=\{1,2\}} \mu \,\hat a_{m,s}^\dagger\hat a_{m,s}    ,
        \label{eq:Hamiltonian}
 \end{align}
where  $\hat a_{m,s}$   with $s=1,2$ are bosonic annihilation operators,  
 $ \nu (t)  \equiv \nu_0 +\nu_1 \cos(\Omega t)$, $ \nu' (t)  \equiv \nu_0' +\nu_1' \cos(\Omega t)$ and $\mu$ the chemical potential. The system is sketched in Fig.~\ref{fig:system}(a). For $\nu_1=\nu_1'=0$ the first line
 resembles the famous Su-Schrieffer-Heeger model which exhibits a topological phase transition
for $\nu_0=\nu'_0$ \cite{Heeger1988}.
 Due to the non-particle-conserving terms in the second line the bosonic modes can exhibit
  dynamical instabilities with exponentially growing bosonic occupation.

After a transformation into the momentum space, the Bogoliubov
Hamiltonian reads
 \begin{align}
 \hat H^{(B)}(t)= \frac 12 \sum_k \left(  \mathbf{\hat a}^\dagger_{k}, \mathbf{ \hat a}_{-k}   \right) 
\mathbf {H}_{k}(t)
 \left(
 \begin{array}{c}
                                                                                                      \mathbf{\hat a}_{k} \\
                                                                                                     \mathbf{\hat a}_{-k}^\dagger
  \end{array}
  \right)       .
  \label{eq:BogoliubovHamiltonian}
   \end{align}
   The symbol $\mathbf{\hat a}_k^\dagger = \left( \hat a^\dagger_{k,1},\hat  a^\dagger_{k,2} \right)$  denotes a vector of  bosonic creation
    operators, and
  \begin{align}
\mathbf {H}_{k}(t)&\equiv \mathbbm 1 \otimes \vec h(k,t) \vec \sigma   - \mu  \mathbbm 1 \otimes \mathbbm 1  + g \sigma_x \otimes \mathbbm 1   ,
  \label{eq:MomentumHamiltonian}
\end{align}
where $\vec \sigma= \left( \sigma_x,\sigma_y\right)$ is a vector of Pauli matrices and we defined a pseudo
magnetic-field vector with components $h_x(k,t)= -\nu(t)-\nu'(t) \cos k\equiv h_{x,0}(k)+h_{x,1}(k)\cos \Omega t $ and $h_y(k,t)= -\nu'(t) \sin k\equiv h_{y,0}(k)+h_{y,1}(k)\cos \Omega t$.

Importantly, the Bogoliubov Hamiltonian fulfills a generalized chiral symmetry at all times $t$. This is defined by
\begin{align}
 \boldsymbol \Sigma  &\left[ \mathbf {H}_{k}(t)+ \mu \mathbbm 1 \otimes \mathbbm 1  - g \sigma_x \otimes \mathbbm 1  \right] \boldsymbol \Sigma  \nonumber  \\ &= -\left[ \mathbf {H}_{k}(t)+ \mu \mathbbm 1 \otimes \mathbbm 1  - g \sigma_x \otimes \mathbbm 1  \right],
 \label{eq:ChiralSymmetry}
\end{align}
where $\boldsymbol \Sigma= \sigma_z \otimes \sigma_z$. Accordingly, the Hamiltonian corresponds to the topological class BDI according to 
the Altland-Zirnbauer classification~\cite{Altland1997}.
The band structure of the undriven and non-interacting system exhibits two bands. They are described by
a topological quantum
number  given by
\begin{equation}
 W  =  \frac 1{2\pi i} \int_{-\pi}^{\pi} \frac d {dk} \ln \left[ h_x(k) + i h_y(k)\right],
\label{eq:defWindingNumb}
\end{equation}
which counts how often the  vector $\vec h(k)$ winds around $\vec h(k)=0$. 
Consequently, $ W$ can only change, if there is a degeneracy as this
is related to $\vec h(k)=0$. By definition, the winding number describes
translational invariant systems. However, there is an important consequence for
finite-sized systems with boundaries. There are spatially confined states close
to the boundary with energy located within the band gap. The number of
these states equals $ W$~\cite{Hasan2010,Gurarie2011}. 
We now show how the physics is modified  in the presence of periodic driving and interactions.

\textit{Floquet-Bogoliubov Theory.}
\label{sec:FloquetBogTheory}
As  regular Bogoliubov excitation energies of an undriven system, the Floquet-Bogoliubov quasienergies for bosons
reveal the stability of  a system. It is stable, if all quasienergies are real-valued and unstable if one or 
more have a finite imaginary part. 
They can be obtained in analogy to the undriven case ~\cite{Goren2007,Kawaguchi2012,Colpa1978,Salerno2016,Arnold1978}: first, we have to solve the differential
equation
\begin{equation}
i \frac d{dt} \mathbf U(t) =  \boldsymbol{\sigma_z}   \mathbf {H}(t)     \mathbf U(t),   \qquad \qquad   \mathbf U(0) = \mathbf 1,
\label{eq:timeEvolutionOperator}
\end{equation} 
where $\boldsymbol{\sigma_z}= \sigma_z \otimes \mathbbm 1$ emerges as the Bogoliubov Hamiltonian couples bosonic creation and annihilation operators, whose equation of motion
differ by a minus sign.
The  matrix $ \mathbf U(2\pi/ \Omega)$ is the Floquet operator and its eigenvalues and eigenstates fulfill

\begin{equation}
   \mathbf U(2\pi/ \Omega) \left|\Psi_{i} \right> =  e^{- i \frac{2\pi}{\Omega} \epsilon_{i}  } \left|\Psi_{i} \right>,
\end{equation}
where $\epsilon_i$  denotes the Bogoliubov quasienergies and $\left|\Psi_i \right>  $ the stroboscopic Floquet states~\cite{Kolovsky2009}. 
There are always $2 d$ Floquet states where $d$ denotes the dimension of the single-particle Hamiltonian.
As  usual quasienergies, the  real part of the Bogoliubov quasienergies can be represented within the window $\left(-\frac \Omega 2,\frac \Omega 2 \right)$.
 A system   is only  stable if the quasienergies for all $k\in (-\pi,\pi)$ are real-valued. In this
 case we denote the system to be globally stable.

Additionally, we introduce the concept of strong stability according to Refs.~\cite{Arnold1989,Starzhinskii1975}. A Floquet state with $\text{Im} \,\epsilon_i=0$ is denoted to be \textit{strongly} stable, if a small perturbation of the system does not result in a finite imaginary part $\text{Im}\, \epsilon_i\neq0$. If a state is strongly stable, then it can be normalized as $C_i\equiv \left<\Psi_i\right|  \boldsymbol{\sigma_z}  \left|\Psi_i \right>=\pm1$~\cite{Kawaguchi2012,Starzhinskii1975}. For every  state with $\epsilon_{i}$ there
is a corresponding state with  $\epsilon_{i'}=-\epsilon_{i}$.
If the states are normalizable, then $C_i=-C_{i'}$.
A system is considered to be strongly stable, if all Floquet states are strongly stable.

\begin{figure}[t]
  \centering
  \includegraphics[width=1\linewidth]{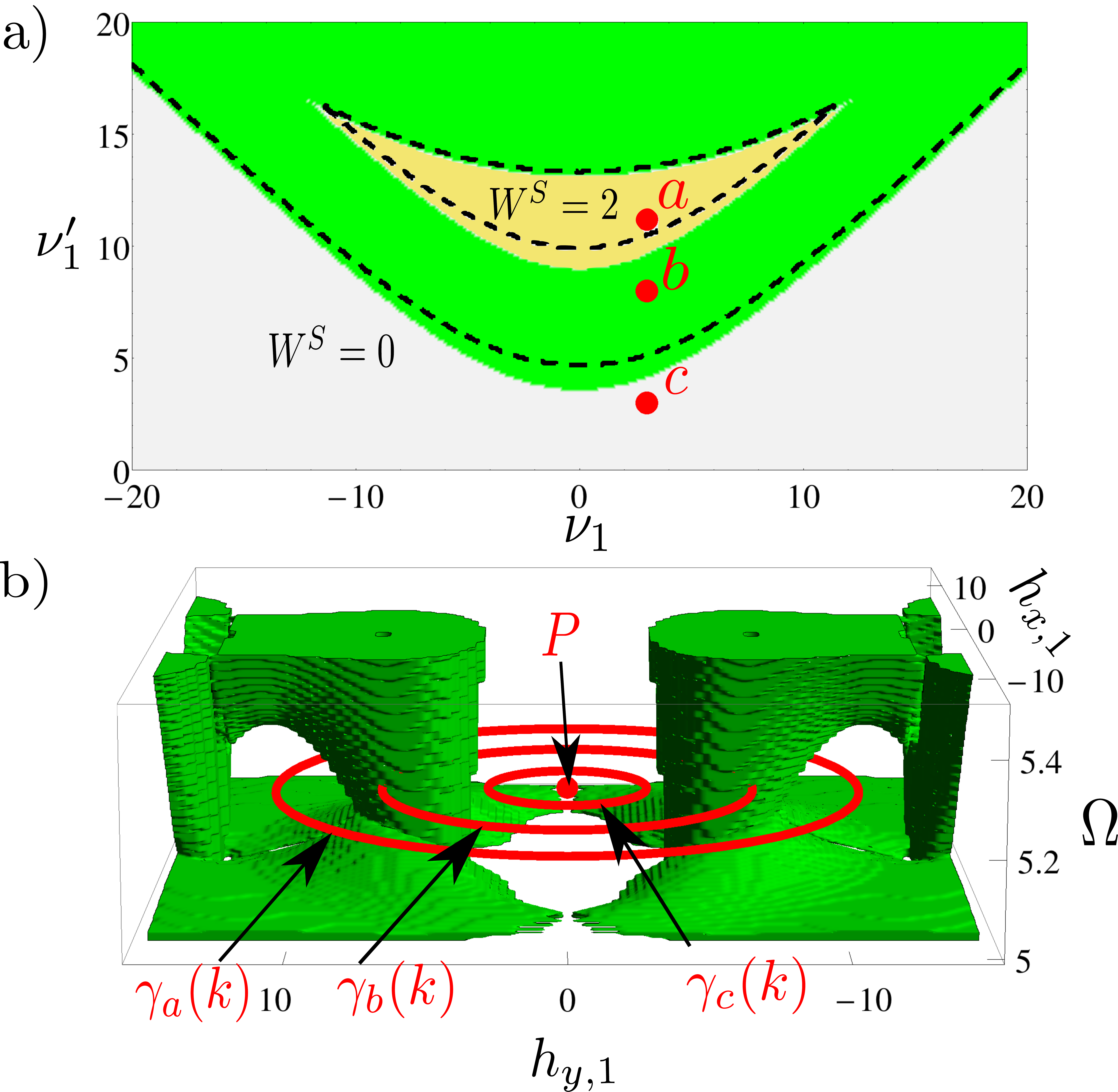}
  \caption{(Color online) \textbf{(a) }Topological phase diagram. The parameters are as in Fig.~\ref{fig:system}(b). In the green  areas, the system is not globally stable, so the topological invariant $W^S$ in Eq.~\eqref{eq:symplecticPolarization}
  is not defined there. In the gray and yellow areas, we find that $W^S=0$ and $W^S=2$, respectively. The topological phases are separated by instability areas (green). Dashed lines are obtained by using an effective Hamiltonian~\cite{Supplementals}. \textbf{(b)} Stability diagram as a function of $h_{x,1}$ and $h_{y,1}$ corresponding to (a). The parameters of the curves $\gamma_{a,b,c}$ 
  are depicted in (a) by the points $p=a,b,c$. The parameters $p=a$  and $p=b$ correspond to Fig.~\ref{fig:system}(b) and (c). Curve $\gamma_c$ can be contracted to  point $P$, so that it is topologically trivial according to the explanations in the main text.
  }
  \label{fig:phaseDiagram}
  \end{figure}

Two Bogoliubov quasienergy dispersions are depicted in Fig.~\ref{fig:system}(b) and (c), where we take the state with $C_{k,i}=1$ if it is normalizable.
In  panel (c), we recognize momenta $k$ with $\text{Im } \epsilon_{k,i} \neq 0$, leading to a dynamical instability with
bosonic occupations growing exponentially in time~\cite{Kawaguchi2012,Goren2007}, which we analyze 
in Fig.~\ref{fig:phaseDiagram}(a), where the system is not globally stable in the green areas.

Following Ref.~\cite{Arnold1989,Starzhinskii1975}, one finds that if 
the quasienergies of two strongly stable states with $C_i \neq C_j$ merge by varying system parameters, thus they become
\begin{equation}
  \text{Re } \epsilon_{i} =  \text{Re } \epsilon_{j} \mod \Omega,
  \label{eq:ResonanceCondition}
\end{equation}
then the states are not strongly stable. Moreover, even when getting unstable, the states $i,j$ still fulfill Eq.~\eqref{eq:ResonanceCondition} which thus constitutes a necessary instability condition.
Consequently, if the quasienergy of an originally strongly stable state  gets  $\text{Re } \epsilon_{i} = 0, \Omega/2 $, it is not strongly stable.   We relate this to 
topology in the following.

 \textit{Topology.} 
 Motivated by~\cite{Shindou2013,Engelhardt2015}, we define a topological invariant  generalizing  Eq.~\eqref{eq:defWindingNumb}  for driven bosonic Bogoliubov systems. 
 \begin{equation}
  W^{S} \equiv \frac 1 {\pi i} \sum_{i\in \mathcal S} \int_{-\pi}^{\pi} dk \left< \Psi_{k,i} \right|\boldsymbol{ \sigma}_z \frac{d}{dk} \left|\Psi_{k,i} \right>,
  \label{eq:symplecticPolarization}
 \end{equation}
 where  $\mathcal S=\left \lbrace i \left|\right. 0< \epsilon_{k,i}< \Omega/2 \land C_{k,i}=1 \text{ for all } k \right \rbrace$ is the set
 of all positive normalizable quasienergies. We note that  $ W^{S}$ is only defined for globally strongly stable systems.
 
  As our Hamiltonian fulfills a generalized chiral symmetry, the topological invariant is integer valued, i.e., $W^{S} \in \mathbbm Z $. 
 It predicts midgap states energetically located close to $\epsilon_i=0,\Omega/2$, thus, close to the instability condition~\eqref{eq:ResonanceCondition}.
The number of midgap states at each boundary equals $W^S$.

The topological phase diagram is depicted in Fig.~\ref{fig:phaseDiagram}(a). There we find a phase with $W^S=0$ (gray) and one with $W^S=2$ (yellow).
Interestingly, these two phases  are separated by instability regions which is a general feature  in driven bosonic
systems, cf. below.

\textit{Instability-topology relation.}
We are now in a position to establish a general relation between  topology and   instability of a bosonic system under ac-driving: The topological invariant, Eq.~\eqref{eq:symplecticPolarization}, can only change  by a smooth variation of system parameters $p(t)$ with $t\in\left[0,1\right]$, if the system is not globally strongly stable for at least one $t=t_0$.  

The relation is a direct consequence of the instability condition Eq.~\eqref{eq:ResonanceCondition}.
When we only perform parameter variations  so that the system is globally strongly stable,  Eq.\eqref{eq:ResonanceCondition} is never fulfilled and   $\mathcal S$ remains unchanged. Moreover, the bands  $i\in \mathcal S$ are not in contact with the bands  $i\neq \mathcal S$ so that topological invariant can not change.

 By definition, in the vicinity of stable but not strongly stable parameters there are always unstable parameters. Consequently, the topological phases in Fig.~\ref{fig:phaseDiagram}(a) are separated by unstable regions.
 
 We emphasize that this relation is valid for systems of arbitrary dimensions. One only has  to replace $W^S  $ in  Eq.~\eqref{eq:symplecticPolarization} by a higher-dimensional topological invariant.
 
 To elucidate this relation, we investigate the 
 stability of the Hamiltonian \eqref{eq:BogoliubovHamiltonian} as a function of  $h_{\eta,1}$ with  $\eta=x,y$ and $\Omega$. 
 The result  is depicted in Fig.~\ref{fig:phaseDiagram}(b), where we depict
 unstable parameters in green.
 A set of system parameters $p$ specifies a curve as a function of monentum k in the stability diagram due to the parametrization of $h_{\eta,1}$ with $\eta=x,y$ below Eq.(3). 
 For instance, the points  $p=a,b,c$  depicted in Fig.~\ref{fig:phaseDiagram}(a) correspond to curves $\gamma_p(k)$
 in the stability diagram in (b).
 The curves $\gamma_a $ and $\gamma_c$ do not traverse any unstable areas, so they are globally stable,
 while $\gamma_b$ is not.

 If one can smoothly contract a globally strongly stable $\gamma(k)$ while clearly avoiding areas of unstable parameters, then $  W _\gamma^{S}=0$. In Fig.~\ref{fig:phaseDiagram}(b), $\gamma_c(k)$ can be trivially contracted to $P$ so that it has a trivial topology.
 By contrast,  the deformation of $\gamma_a(k)$ onto a stable point  in Fig.~\ref{fig:phaseDiagram}(b),
 is only possible by traversing the two unstable green regions. 
 For this reason, it is a canditate for a non-trivial topology with  $ W_\gamma^{S}\neq 0$.
  We thus have a one-to-one correspondence
 between \textit{points} in the topological phase diagram Fig.~\ref{fig:phaseDiagram}(a), and (non) contractible \textit{curves} in the stability diagram in  Fig.~\ref{fig:phaseDiagram}(b).
 
 Moreover, the stability diagram assists to find parameters corresponding to a stable and topological-nontrivial system. We only have to calculate $  W^{S}$ for non-contractible curves which can not be transformed to each other in a stable way.

The black lines in Figs.~\ref{fig:system}(b),(c) and \ref{fig:phaseDiagram}(a) depict the calculations using a time-independent effective Hamiltonian~\cite{Supplementals}. 
 For its derivation, we generalized the procedure of Ref.~\cite{Benito2014} to  Bogoliubov Hamiltonians which, to our knowledge, has not been done before. As the effective Hamiltonian resembles the  features of the spectrum, it is an appropriate tool to calculate  stability diagrams which enable an efficient search for parameters of  stable and topologically non-trivial systems.

\begin{figure}[t]
  \centering
  \includegraphics[width=1\linewidth]{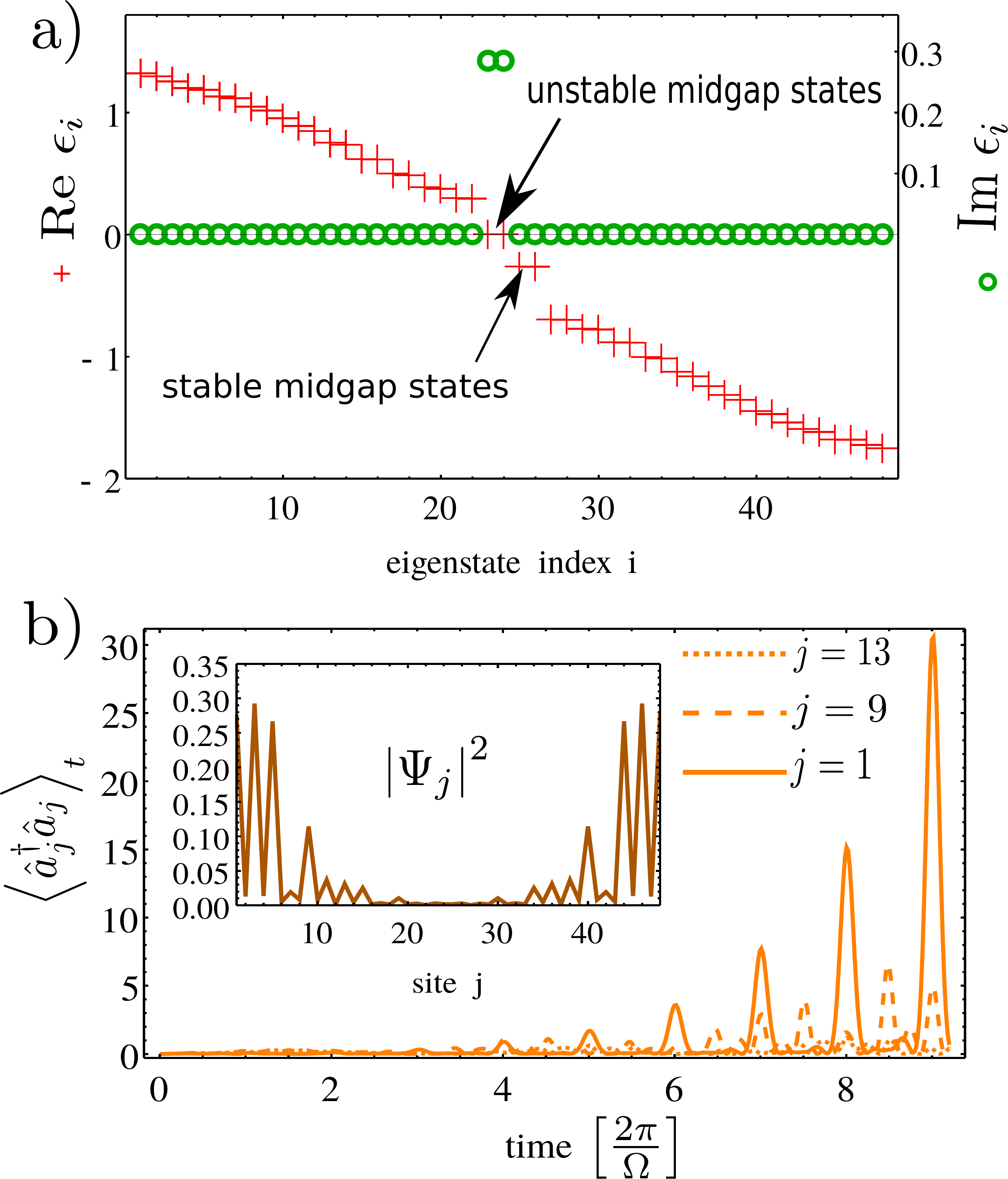}
  \caption{(Color online) \textbf{(a)} Quasienergy spectrum for a finite-sized system with boundaries corresponding to  $\gamma_a$. 
  There are four midgap states located near $\text{Re }\epsilon_i=0$ which we mark with arrows. 
  The imaginary part of two of the midgap states
       is finite, which renders these states unstable. States not marked by the arrows  extend within the bulk and are stable. 
     \textbf{(b)} Time evolution of the occupation of the sites $j=1,9,13$. The initial state is the vacuum state. In the inset we
     depict one of the unstable midgap states which are responsible for the exponential growth as seen in  (b).  
     It is strongly confined close to the boundaries.
       Note that the site number is $j=2m +s$, with $m,s$ defined in Eq.~\eqref{eq:Hamiltonian}.
  }
  \label{fig:spectrum}
  \end{figure}

 \textit{ Topological instabilities.}
  \label{sec:EdgeStateInstability}
Due to its definition, $  W ^{S}$   predicts midgap states energetically located close to $\epsilon_i=0,\Omega/2$, thus, close to the instability condition~\eqref{eq:ResonanceCondition}.
 In Fig.~\ref{fig:spectrum}(a), we depict the numerical quasienergy spectrum for a finite-sized system with boundaries   with parameters given by $p=a$ in Fig.~\ref{fig:phaseDiagram}(a).
 There are four midgap states with $\text{Re } \epsilon_i \approx 0 $ which we mark with arrows. Their wave functions are strongly
 confined to the boundaries. The wave function of one of them is depicted in the inset of panel (b). 
  The imaginary part of the quasienergies of two midgap states is finite so they are dynamically unstable, which gives rise to an exponential growth of
  the occupation as a function of time.
  The imaginary part of the quasienergies of the other two midgap states is zero so
 they are stable. 
 The states not marked by an arrow are bulk modes and are stable.

In principle,  one can render all four midgap states to be unstable by slightly adjusting the system parameters as $\text{Re } \epsilon_i \approx 0 $.
 However, we did not
 find parameters where all four midgap states are unstable without destabilizing the bulk modes.

The initial state of the time evolution in Fig.~\ref{fig:spectrum}(b) is the vacuum state defined by
$\hat a_{j}\left| \text{vac}\right>=0$~\cite{Barnett2013,Furukawa2015,Galilo2015}. We recognize, that
the occupation grows exponentially on sites close to the boundary, while it remains small within the bulk.
As the instabilities are generated by the unstable midgap states, they can be thus considered to be topologically protected instabilities.

%


To conclude, 
we found that topological phases are separated by regions of instability in ac-driven systems which we illustrated using stability diagrams. We recall that this finding is valid for systems of arbitrary dimension.
We used this   to selectively generate dynamical instabilities which are strongly localized close to the boundaries. To this end, we employed localized midgap states whose occupation grows exponentially in time.
Recently, stability and the onset of chaos have been experimentally explored in a periodically-driven 
two-mode Bose-Einstein condensate \cite{Tomkovivc2015}. We assume that the spatially extended system investigated here is also a
candidate for  such an experimental investigation.

\textit{Acknowledgments}
The authors gratefully acknowledge financial support
from the DFG Grants No. BR 1528/7, No. BR 1528/8,
No. BR 1528/9, No. SFB 910 and No. GRK 1558. 
 This work was supported by the Spanish Ministry
through Grant No. MAT2014-58241-P.

\bibliographystyle{apsrev4-1}
 \bibliography{topology} 

 \begin{widetext}

\begin{center}
   \huge \textbf{Supplementary Information}
   \vspace{0.7cm}
   
   \Large \textbf{Derivation of the effective Hamiltonian}
\end{center}

In order to efficiently determine the full parameter range where unstable midgap states can be expected,
we derive a time-independent effective Hamiltonian. We generalize the procedure of Ref.~\cite{Benito2014} to  Bogoliubov Hamiltonians which, to our knowledge,
has not been done before.

The equation of motion  can be written as
\begin{equation}
   i \frac d {dt} \left| \Psi_k \right> = \boldsymbol{\sigma_z}  \left(\mathbf H_{k,0} + \mathbf H_{k,1}\cos \Omega t \right)   \left| \Psi_k \right>,
\end{equation}
where
\begin{align}
    \mathbf H_{k,0} &= \mathbbm 1 \otimes \left[ -\mu  \mathbbm 1+ h_{x,0} (k)\sigma_x + h_{y,0} (k)\sigma_y  \right] + g\, \sigma_x \otimes \mathbbm 1\\
    \mathbf H_{k,1} &=  \mathbbm 1 \otimes \left[  h_{x,1} (k)\sigma_x + h_{y,1}(k) \sigma_y  \right] .
\end{align}
In the following, we suppress the dependence of momentum $k$ in the coefficients $h_{x,0},h_{y,0},h_{x,1},h_{y,1}$ for a notational reason.
We transform this equation into an interaction picture. This is defined by the unitary transformation
\begin{equation}
\mathbf U_{k,\alpha,\beta}(t) = \exp \left[ - i \boldsymbol{\sigma_z}  \frac{\mathbf H_{k,1} }{\sqrt[4]{\text{det}\mathbf H_{k,1} }}\theta_\alpha (t), \right] \exp \left[ -i\boldsymbol{\sigma_z} \frac{\beta \Omega t}2  \right] ,
\end{equation}
where 
\begin{equation}
\theta_{k,\alpha} (t )= \sqrt[4]{\text{det}\mathbf H_{k,1} }  \frac{\sin \Omega t}{\Omega} + \frac{\alpha \Omega}{2} t .
\end{equation}
The parameters $\alpha,\beta$  are integers.  
In the interaction picture, the equation of motion reads
\begin{equation}
   i \frac d {dt} \left| \Psi_{k,I} \right> = \mathbf U_{k,\alpha,\beta}^\dagger (t)   \boldsymbol{\sigma_z} \left( \mathbf H_{k,0} - \alpha   \frac{\Omega} {2} \frac{\mathbf H_{k,1} }{\sqrt[4]{  \text{det}\mathbf H_{k,1} } } - \frac{\beta \Omega}2  \right) \mathbf U_{k,\alpha,\beta}(t ) \left| \Psi_{k,I} \right>.
\end{equation}
The integer $\alpha$ can be chosen to effectively reduce the gap between the two bands and the integer $\beta$ shifts the chemical potential $\mu$ which can be understood later by considering Eqs.~\eqref{eq:reducedGap} and \eqref{eq:ReducedChemPot}. 
 First, we calculate an explicit expression for $ \mathbf  U_{k,\alpha,\beta}(t)$. 
Defining $\phi_k \equiv \text{Im }  \ln\left( h_{x,1} + i h_{y,1}\right)$,
we obtain
\begin{align}
\mathbf  U_{k,\alpha,\beta}(t) & =  \left(
              \begin{array}{cccc}
                          U_{k,\alpha,\beta} \; e^{- i  \frac{\beta \Omega}2 t   } & 0 \\
                            0                       &  U_{k,\alpha,\beta}^\dagger \;  e^{ i   \frac{\beta \Omega}2 t  }
              \end{array}
     \right) ,
\end{align}
where
\begin{align}
  U_{k,\alpha,\beta}(t) &=     \mathbbm 1 \cos \theta_{k,\alpha}(t) + i \sigma_x   \sin \theta_{k,\alpha}(t) \cos \phi_k + i \sigma_y  \sin \theta_{k,\alpha}(t) \sin \phi_k   .
\end{align}
For a notational reason we define
\begin{align}
    h_{\eta}^\alpha&= h_{\eta,0}- \frac{\alpha \Omega}{2}  \frac{h_{\eta,1}}{\sqrt[4]{\text{det}\mathbf H_{k,1} }} \qquad \text{with} \quad \eta=x,y,
    \label{eq:reducedGap}
\end{align}
where we again suppress the argument $k$. 
We evaluate the matrix product
\begin{align}
   & U_{k, \alpha,\beta }^ \dagger   \left[ h_{x}^\alpha \sigma_x  +    h_{y}^\alpha \sigma_y \right]  U_{ k,\alpha ,\beta} = \nonumber  \\
    &=  \sigma_x  \left[  \cos^2\theta_{k,\alpha}(t)  h_{x}^\alpha  +  \sin^2 \theta_{k,\alpha}(t) \left( h_{x}^\alpha \cos^2 \phi_k- h_{x}^\alpha \sin^2 \phi_k + 2 h_{y}^\alpha \cos\phi_k \sin \phi_k \right)   \right] \nonumber  \\
                &+ \sigma_y  \left[  \cos^2\theta_{k,\alpha}(t) h_{y}^\alpha  + \sin^2 \theta_{k,\alpha}(t) \left( h_{y}^\alpha \sin^2 \phi_k- h_{y}^\alpha \cos^2 \phi_k   + 2 h_{x}^\alpha \cos\phi_k \sin \phi_k \right)   \right]  \nonumber\\
                &+  (h_{y}^\alpha\cos \phi_{k} -h_{x}^\alpha\sin \phi_{k}) 2 \sin\theta_{k,\alpha}(t) \cos\theta_{k,\alpha}(t)  \sigma_z  \nonumber \\
                &\equiv \tilde h_{x}^\alpha(t) \sigma_x  +   \tilde h_{y}^\alpha(t) \sigma_y+  \tilde h_{z}^\alpha(t) \sigma_z.
\end{align}
Finally we apply a rotating-wave approximation by a time average of the coefficients. In doing so we use that
\begin{align}
&\frac{1}{2\pi/\Omega} \int_0^{2\pi/ \Omega} \cos^2 \theta_{k,\alpha}(t) = \frac 12 \left[1+ \mathcal J_\alpha \left( \frac {2  \sqrt[4]{\text{det}\mathbf H_{k,1} }  }{\Omega} \right) \right],\\
&\frac{1}{2\pi/\Omega}\int_0^{2\pi/ \Omega} \sin^2 \theta_{k,\alpha}(t) = \frac 12 \left[1- \mathcal J_\alpha \left( \frac {2  \sqrt[4]{\text{det}\mathbf H_{k,1} } }{\Omega} \right) \right] ,\\
&\frac{1}{2\pi/\Omega}\int_0^{2\pi/ \Omega} \sin \theta_{k,\alpha}(t) \cos \theta_{k,\alpha} (t) = 0 ,
\end{align}
where $\mathcal J_\alpha(x)$ denotes the Bessel function of order $\alpha$.
Analogously, we treat the term
\begin{align}
 & U_{k, \alpha,\beta }^{ \dagger} (t) \; g \mathbbm 1 \;  U_{k, \alpha,\beta }^\dagger(t).
\end{align}
Finally, we obtain 
\begin{align}
\mathbf {H}_{\text{eff},k}&= \mathbbm 1 \otimes \vec{h}^\alpha_{\text{eff}}(k) \; \vec \sigma   -   \mathbbm 1 \otimes \mu^\beta_{\text{eff}}\; \mathbbm 1   \nonumber \\
        &+ \sigma_x \otimes \vec G^{ \alpha, \beta}_{\text{eff}}(k)\; \vec \sigma +  \sigma_x \otimes g^{\alpha, \beta}_{\text{eff}}(k) \; \mathbbm 1   ,
  \label{eq:effHamiltonian}
\end{align}
 with the coefficients
\begin{align}
 h_{\text{eff},x}^\alpha (k) &=  h_{x}^\alpha \left[ f_k^+\left( \frac{h_{y}^\alpha}{h_{x}^\alpha}\right) + f_k^-\left( \frac{h_{y}^\alpha}{h_{x}^\alpha}\right) \mathcal J_\alpha \left( \frac {2  \sqrt[4]{\text{det}\mathbf H_{k,1} } }{\Omega} \right)  \right], \\
   h_{\text{eff},y}^\alpha (k)&=  h_{y}^\alpha \left[ f_k^-\left(- \frac{h_{x}^\alpha}{h_{y}^\alpha}\right) + f_k^+\left( -\frac{h_{x}^\alpha}{h_{y}^\alpha}\right) \mathcal J_\alpha \left( \frac {2  \sqrt[4]{\text{det}\mathbf H_{k,1} } }{\Omega} \right)  \right] , \\
\mu^\beta_{\text{eff}} &= \mu - \beta \frac\Omega 2, \label{eq:ReducedChemPot}\\
g^{\alpha, \beta}_{\text{eff}} (k)&=  \frac g 2 \left[  J_{-\beta-\alpha} \left( \frac {2  \sqrt[4]{\text{det}\mathbf H_{k,1} }  }{\Omega} \right) +J_{\beta-\alpha} \left( \frac { 2  \sqrt[4]{\text{det}\mathbf H_{k,1} }  }{\Omega} \right)   \right] ,
\\
 G^{\alpha, \beta}_{\text{eff},x}(k) &= \frac g 2 \cos \phi \left[  J_{-\beta-\alpha} \left( \frac {2  \sqrt[4]{\text{det}\mathbf H_{k,1} }  }{\Omega} \right) -J_{\beta-\alpha} \left( \frac { 2  \sqrt[4]{\text{det}\mathbf H_{k,1} }  }{\Omega} \right)   \right] , \\
 G^{\alpha, \beta}_{\text{eff},y}(k)  &= \frac g 2 \sin \phi \left[  J_{-\beta-\alpha} \left( \frac {2  \sqrt[4]{\text{det}\mathbf H_{k,1} }  }{\Omega} \right) -J_{\beta-\alpha} \left( \frac { 2  \sqrt[4]{\text{det}\mathbf H_{k,1} }  }{\Omega} \right)   \right],
\end{align}
where we have defined
\begin{align}
2 f_k^{\pm}(x) &\equiv 1\pm \cos^2 \phi_k \mp \sin^2 \phi_k \pm 2 x \sin \phi_k\cos \phi_k .
\end{align}
This expression is valid if $h_{x}^\alpha,h_{y}^\alpha, \mu^\beta_{\text{eff}},g \ll \Omega$.
In order to fulfill $\mu^\beta_{\text{eff}} \ll \Omega$, we choose $\beta$ so that $\left|\mu^\beta_{\text{eff}} \right| $ is minimal, thus  $\left|\mu^\beta_{\text{eff}} \right| <\Omega/4$. In a similar way we have to choose $\alpha$, so that the modulus of $h_{x}^\alpha,h_{y}^\alpha$ defined in Eq.~\eqref{eq:reducedGap} are as small as possible.
The  effective Hamiltonian  still fulfills a generalized chiral symmetry relation~(4) in the letter, independently for
all momenta $k$.
The Hamiltonian Eq.~\eqref{eq:effHamiltonian} can be diagonalized so that we obtain the energies
\begin{align}
\epsilon_{\pm} &= \pm \sqrt{ \left|\vec{h}_{\text{eff}}^\alpha\right|^2 + \left( \mu^{\beta}_{\text{eff}}\right)^2 -\left(g^{ \beta}_{\text{eff}}\right)^2  -\left(\tilde g^{ \beta}_1\right)^2 \pm 2 \sqrt A}.\nonumber \\
A &= - \tilde g^{ \beta}_1 \left|\vec{h}_{\text{eff}}^\alpha\right|     \left(\tilde g^{ \beta}_1 \left|\vec{h}_{\text{eff}}^\alpha\right|\sin^2 \delta \phi +   \tilde g^{ \beta}_{\text{eff}} \mu^{\beta} \sin^2 \frac{\delta \phi}2{} \right) \nonumber 
 +\left(g^{ \beta}_{\text{eff}}  \tilde g^{ \beta}_1  +  \left|\vec{h}_{\text{eff}}^\alpha\right| \mu^{\beta}   \right)^2   .
\label{eq:effQuasienergies}
\end{align}
where $\delta\phi  $ is the angle between $ \vec G^{ \beta}_{\text{eff}}$ and $\vec{h}_{\text{eff}}$ . The $\epsilon_{\pm}$ resemble the actual quasienergies of the system modulo $\Omega/2$ as a consequence of ~Eq.~\eqref{eq:ReducedChemPot} shifting the chemical potential, and the folding of the quasienergies into the window $(-\Omega/2,\Omega/2)$.
We compare the analytical calculated $\epsilon_{\pm}$ and the numerical quasienergies in Fig.~1(b) and (c), where we choose $\alpha=0$ and $\beta=-2$ so that the condition of validity is fulfilled. The effective Hamiltonian
reproduces all features of our numerics, especially the instabilities, although we work with a mediate frequency $\Omega$.
Corresponding to   the actual quasienergies, the system is stable if all $\epsilon_{\pm}$ are real valued. In Fig.~2(a) we use the $\epsilon_{\pm}$ to calculate the regions of global stability. The result is depicted with the black dashed lines.

 \end{widetext}

\end{document}